# Design and Development of a Web Platform for Blood Donation Management

Fatima Zulfiqar Ali[1] and Atrooba Ilyas[1,2]

[1]Department of Computer Science and Information Technology, Virtual University of Pakistan,
P.O. Box: 05450, Lahore, Pakistan
[1]bc200412015@vu.edu.pk, [2]atrooba_ilyas488@gmail.com

**Abstract.** Blood donation is a critical component of healthcare, yet locating suitable donors in emergencies often presents significant challenges. This paper presents the design and development of a Blood Donation Web Platform, a web-based system that connects patients, donors, and administrators within a centralized digital space. The platform allows interested donors to register their personal information, including blood group, contact details, and availability. Patients can search for donors based on blood group and location, and the system provides a list of nearby donors who are ready to donate. The platform's design was guided by use case, database, class, and sequence diagrams to ensure a well-structured and efficient system architecture. Modern web technologies, including PHP (Laravel framework), HTML, CSS, Bootstrap, and MySQL, supported by XAMPP and Visual Studio Code, were employed to implement a dynamic, interactive, and user-friendly platform. By streamlining donor registration, blood requests, and communication, the proposed system reduces delays and complexities in emergencies, improving timely accessibility of blood and enhancing overall efficiency in blood donation services.

**Keywords:** Blood Donation, Web Platform, Donor Management, Use Case Modeling, Database Design, PHP Development, System Testing

## 1 Introduction

Blood donation is an act of kindness that helps save the lives of people in need by directly supporting patients during surgeries, accidents, cancer treatments, and other medical situations where blood is urgently required [1]. It is a process of collecting blood from a donor to be used for treating others and plays a vital role in supporting the healthcare system[2]. However, one of the biggest challenges faced during emergencies is the difficulty in locating suitable donors quickly and efficiently. Traditional methods such as contacting relatives, friends, or relying on hospital databases are often time-consuming and may not guarantee the availability of the required blood type [1, 3]. In many cases, the delay in finding a compatible donor can put the patient's life at serious risk. Moreover, the lack of a centralized and updated record of donors further complicates the process, especially in critical situations where every second counts [4]. This highlights the urgent need for a reliable and accessible digital solution that can instantly connect patients with potential donors in real time.

To address this challenge, a new and state-of-the-art Blood Donation Web Platform was developed in this research work to provide a centralized digital space that connects patients, donors, and administrators. The platform enables interested donors to register their personal details, including blood group, contact information, and availability. In emergencies, patients can simply enter the required blood group to view a list of nearby donors who are available and willing to donate [5]. Once a donation is made, the donor's information is temporarily hidden from the system for the next three months [6].

In this paper, we developed a state-of-the-art Blood Donation Web Platform using various software tools and design techniques, discussed in Section, to enhance the efficiency, accessibility, and real-time patient-donor connectivity of blood donation services. The platform's design was guided by multiple modeling diagrams to ensure a clear and well-structured system architecture. The use case diagram, discussed in Section 2.3, outlines the system's boundaries, main functionalities, and the interactions between different actors and the platform [7]. Moreover, Section 2.4 discusses the database diagram highlighting the organization of data, including its structure, relationships, and constraints, ensuring efficient storage and retrieval [8]. The class diagram highlights the static structure of the system by showing key concepts and their relationships [9]. Finally, the sequence diagram formalizes the order of object interactions [10]; a detailed explanation appears in Section. Together, these diagrams provide a comprehensive blueprint that ensures the platform is reliable, scalable, and user-friendly.

## 2    Material and Methodology

### 2.1    Software Platform

The implementation of the Blood Donation Web Platform was performed by using several programming tools and technologies, including Hypertext Preprocessor or Personal Home Page (PHP) [11] programming language, as the Laravel framework based on a Model, View, and Controller (MVC) design model**.** Meanwhile, using Hyper Text Markup Language (HTML) with Cascading Style Sheets (CSS) enables the creation of visually engaging and interactive web pages, making the website more appealing and user-friendly for visitors [12]. My Structured Query Language (MySQL) functioned as the database management system for storing and retrieving donor-related information, supported by phpMyAdmin open source software. Moreover, XAMPP software offered a local development environment with Apache, MySQL, PHP, and Perl, whereas *Visual Studio Code* software served as the primary Integrated Development Environment (IDE) due to its lightweight design, flexibility, and strong support for web development.

### 2.2    Development Methodology for the Blood Donation Platform

The development process of the Blood Donation Web Platform followed a step-by-step structured approach that begins with the collection of system requirements from the primary stakeholders or actors, including Donors, Patients, and the Admin. Each actor has a distinct role based on the system requirements, which are clearly specified in the Use Case Model. Based on these requirements, the system design was prepared using the use case, database, class, and sequence diagrams in order to model the workflow and interactions that are described in later Sections 2.3-2.6 of the article. The database schema was then implemented in MySQL to manage user and donation records efficiently. The front-end interface was developed with HTML and CSS, while backend logic and functionalities were coded in PHP. In the final stage, the platform was rigorously tested to verify functionality, ensure accuracy of operations, and provide a seamless user experience.

## 2.3    Use Case Modeling for Web Platform

The Use Case Model, also referred to as the Use Case Diagram (UCD), is created to visually represent the functional requirements of the web-based system, showing how each actor interacts with the platform according to their role. The model comprises four components: Actors, which represent the primary stakeholders; Attributes, which define the responsibilities and actions of each actor; and Relationships, which illustrate how the actors engage with the system's use cases and the last is system boundary, anything within these boundary is the functionality in scope of the system the rectangle around the use cases [13]. Fig. 1 shows the use case diagram for the web-based system, highlighting the interactions of three primary actors.

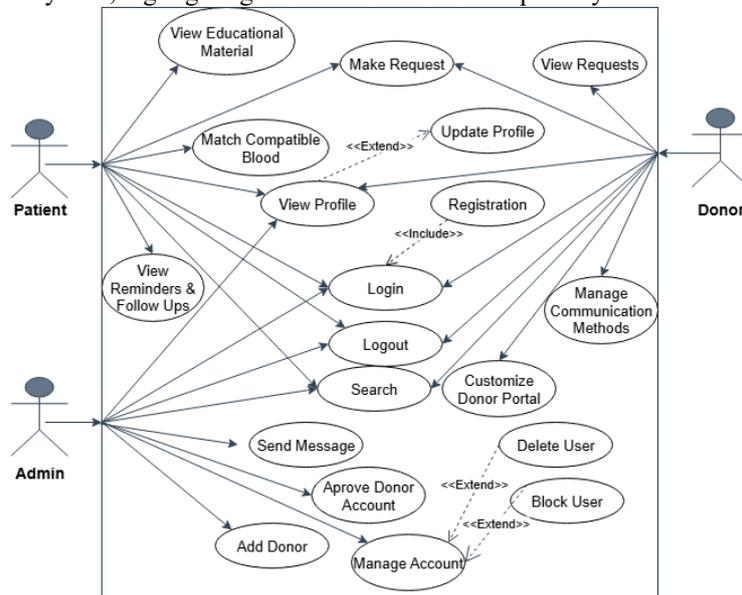

**Fig. 1.** Use Case Diagram of the Blood Donation Web Platform, showing three actors: Admin, Donor, and Patient with their distinct attributes and interactions within the system.

Each actor in the system has specific attributes and responsibilities: the Admin manages users and requests; the Donor provides personal information and donates blood, and the Patient searches for or requests blood. The use case diagram also uses <<include>> relationships to represent mandatory actions shared across use cases and <<extend>> relationships to represent optional or conditional behaviors. These elements together ensure that all system functionalities are clearly defined and efficiently executed.

### 2.4 Database Model (Physical Schema) of the web platform

**Fig. 2** illustrates the database model for the web-based platform, which forms the foundation of the web application by providing a structured framework for storing and managing data.

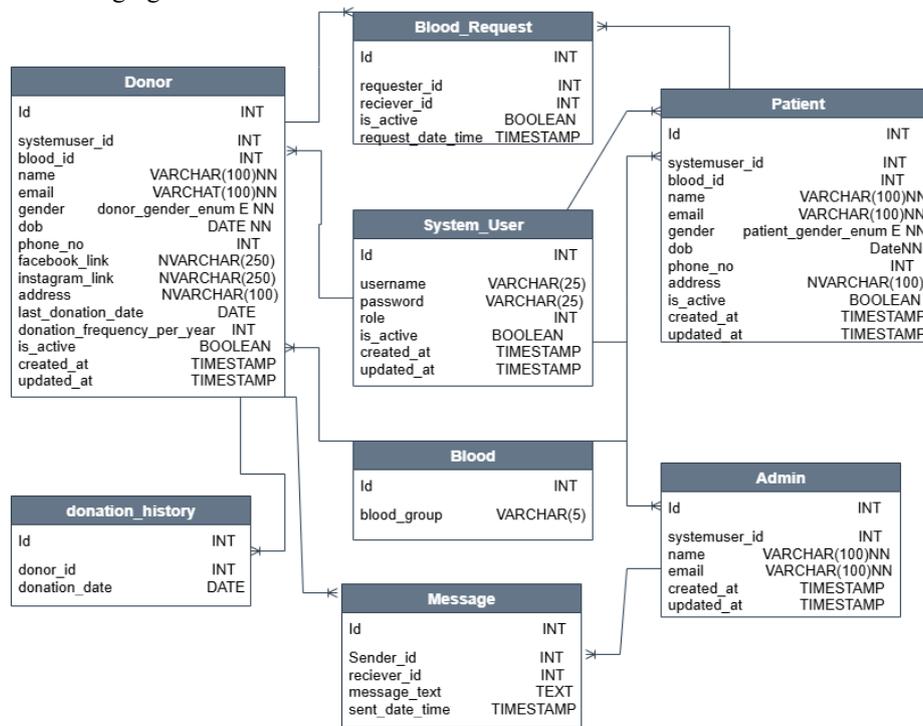

**Fig. 2**. Database Schema of the Blood Donation Web Platform, showing relationships between donors, patients, and administrators

The model is designed and implemented as a relational database in MySQL, which carefully selects the data types to prevent redundancy, maintain data integrity, and enable efficient data retrieval. The user tables are linked to the Donor, Patient, and Admin tables through a one-to-many relationship, indicating that a single user can be associated with multiple donor, patient, or admin records, while each record in these tables belongs to exactly one user. Furthermore, the model defines specific constraints and data types for the users table, as detailed in Table 1.

**Table 1.** Schema of the Users Table for the database diagram

| Field Name | Data Type    | Constraints                |
|------------|--------------|----------------------------|
| **User_id**    | INT          | Primary Key, Auto Increment |
| **Name**       | VARCHAR(100) | NOT NULL                   |
| **Email**      | VARCHAR(100) | Unique, NOT NULL           |
| **Password**   | VARCHAR(25)  | NOT NULL                   |
| **Created_at** | TIMESTAMP    | Default: CURRENT_TIMESTAMP |

User_id is an integer (INT) that uniquely identifies each user and automatically increases for new entries. The Name and Email fields use VARCHAR, which allows storing text up to a specified length (100 characters in this case). The Email field is also marked as Unique to prevent duplicate entries. The Password field uses VARCHAR (25) to store user passwords. The Created_at field uses TIMESTAMP to automatically record the date and time when a record is added. Constraints like NOT NULL ensure that important fields cannot be left empty, while Primary Key and Auto Increment maintain uniqueness and proper indexing of data.

### 2.5   Object Model/Logical Model: Class Diagram

The Class Diagram of the Blood Donation Platform is a type of Unified Modeling Language (UML) used to represent, visualize, and describe the static or Object-Oriented Structure (O-O-S) of the system [14], showcasing the classes, their attributes, methods, and relationships between classes, as illustrated in Fig. 3.

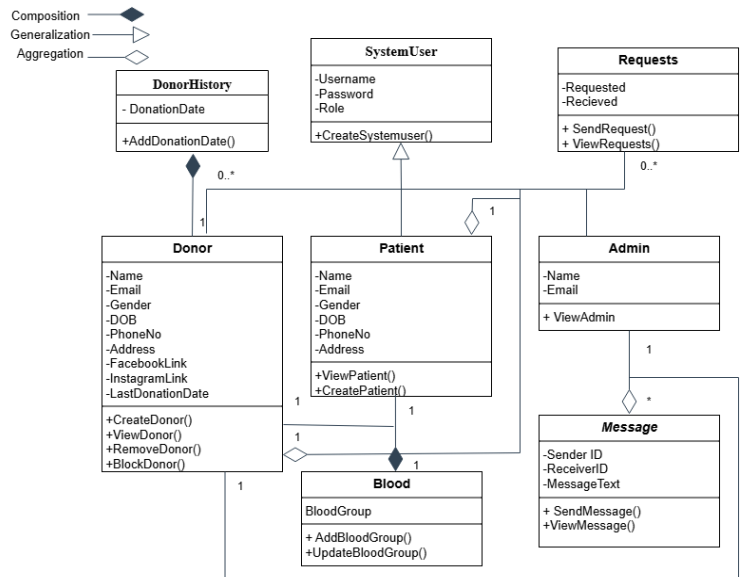

**Fig. 3.** Class diagram of the Blood Donation Web Platform, illustrating the main classes (User, Donor, Patient, and Admin), their attributes, methods, and the relationships among them.

The primary class is User that comprises attributes such as Name, Email, and Password. These attributes are private (−) to protect data, and the methods are kept as public (+) so that such data can be accessed by other parts of the system. Moreover, the specialized classes like Donor, Patient, and Admin inherit from User. The diagram also illustrates relationships: generalization represents the inheritance hierarchy, aggregation shows a 'has-a' relationship without full ownership, and composition represents strong ownership where dependent objects are deleted with the parent.

## 2.6  Sequence Diagram/Interaction Diagram

The Sequence Diagram of a web platform is an interaction diagram that shows how objects or components communicate over time to complete a specific task, illustrating the flow of messages and the chronological order of events. Fig. 4 illustrates the sequence diagrams related to the web platform.

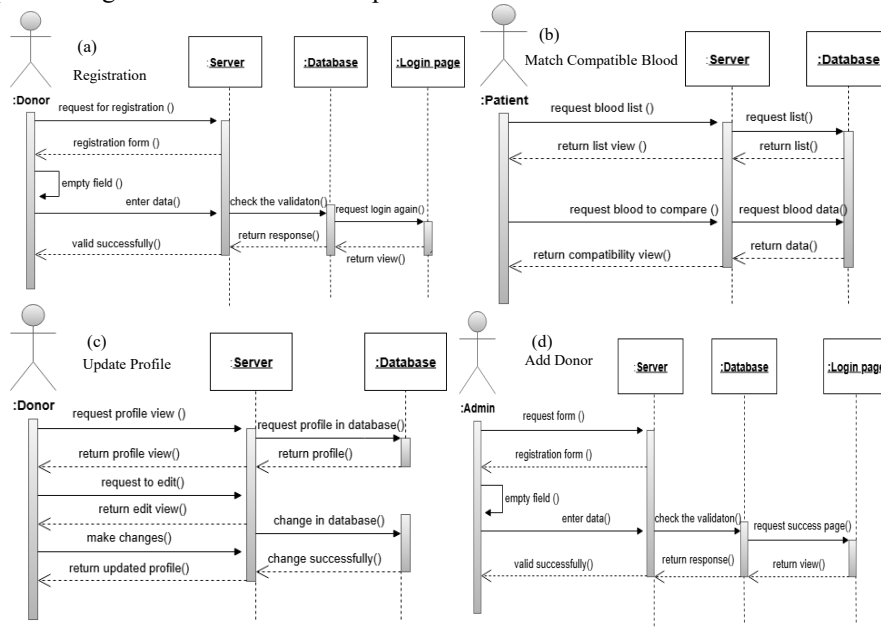

**Fig. 4.** Sequence diagram showing the interactions and message flow among User, Donor, Admin, and System in the blood donation process: (a) Donor registration; (b) Matching compatible blood, (c) Donor profile updates, and (d) Add donor.

The objects or participants are shown as lifelines (vertical dashed lines), like User, Donor, Admin, and System. Arrows between objects or participants show messages or actions. Thin rectangles on a lifeline, called activation bars, show when an object is active. The vertical direction represents the progression of time from top to bottom, making the sequence of interactions clear that can shown in Fig. 4(a). The diagram illustrates the donor registration process, showing how a new donor provides information and is added to the system, and the self-call shows that an empty field means missing information that should have been provided. Whereas Fig. 4(b) depicts the matching of compatible blood for a patient, highlighting the interaction between the donor and patient data. Additionally, Fig. 4(c) shows the donor updating their profile, including the steps for modifying personal information. Finally, Fig. 4(d) demonstrates the admin add donor process, showing the steps involved in registering a new donor into the system.

## 3  Results and Discussions

The Blood Donation Society web application was developed to streamline the process of donor registration, blood requests, and communication between donors, patients, and administrators. The system allows users to securely register and log in, while donors can update their profiles, manage donation availability, and respond to blood requests. Patients can request blood, and the system efficiently matches them with compatible donors, ensuring timely notifications. The admin panel enables administrators to manage users, track requests, and oversee the overall operation of the platform. The application shows improved efficiency, reliability, and ease of use compared to traditional manual systems. While the platform successfully achieves its objectives, future enhancements could include AI-based donor matching, automated reminders, and a mobile-friendly interface to further improve accessibility and responsiveness. Fig. 5 illustrates various components of the web platform.

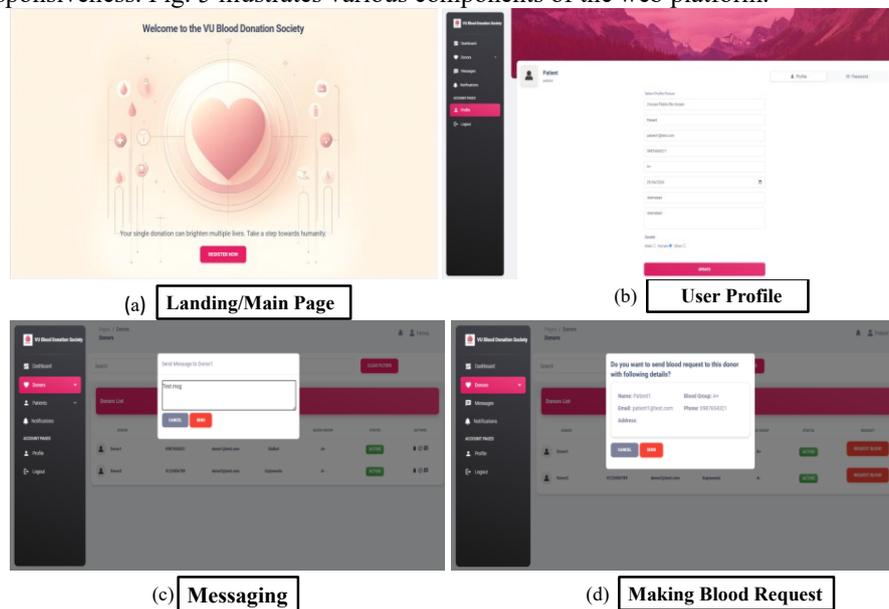

(a) Landing/Main Page  (b) User Profile
(c) Messaging  (d) Making Blood Request

**Fig. 5.** User interfaces of the Blood Donation Society website, showing the landing page, user profile, messaging, and blood request features, including (a) the landing page, (b) the user profile, (c) the messaging system, and (d) the blood request feature.

The website has the following user interfaces: the landing page (Fig. 5(a)), which welcomes users and provides navigation to different sections of the platform; the user profile, as illustrated in Fig. 5(b), where users can view and update their personal information and preferences, the messaging interface (Fig. 5(c)), which allows users to communicate with donors, patients, or administrators; and finally the blood request interface that can be visualized in Fig. 5(d), where patients can submit a request for a specific blood type and quantity, which is then matched with available donors.

### 3.1 Donor Page Design and Backend Implementation

The donor list page, as illustrated in Fig. 6, shows an example of the main Donor Page along with its backend implementation of data retrieval.

**Fig. 6.** Front-end design and layout of the Donor Management Page.

The HTML (Skeleton/Structure) has been used to create the basic structure, including forms, tables, and dropdowns. CSS (Styling) is applied for custom styling, such as colors, fonts, and layout adjustments. Whereas Bootstrap (CSS framework) provides responsive design elements and pre-built components for a professional appearance. PHP (Backend functionality) handles the backend functionality; together, these technologies create a well-structured, styled, and interactive donor page. The donor list page allows users to view details such as Name, Phone, Email, City, Blood Group, and Status. Users can interact with the page to perform actions like editing or deleting records. PHP manages the backend functionality, handling form submissions, fetching donor data from the database, and updating records, making the page fully functional and dynamic.

## 4  Conclusion

The Blood Donation Web Platform developed in this research work provides a centralized and efficient solution for connecting patients in need of blood with registered donors. The use of modern web technologies, including PHP, HTML, CSS, and MySQL, enabled the creation of a dynamic, responsive, and user-friendly system. Furthermore, the modeling tools, such as use case, database, class, and sequence diagrams, ensured clear system architecture, efficient data management, and accurate implementation of functionalities. The platform allows real-time donor searches, simplifies donor registration, and facilitates smooth communication between all stakeholders, addressing the major challenges faced during emergency blood donation situations. Overall, the proposed system demonstrates improved efficiency, reliability, and accessibility in managing blood donations, providing a scalable solution that can be extended to broader healthcare contexts in the future.